\documentclass[aps,preprint,final,prb,12pt]{revtex4}
\usepackage{times}
\begin{document}

\title{
Spin symmetry and spin current of helicity eigenstates
of the Luttinger Hamiltonian
\footnote{The submitted manuscript has been authored by a contractor of the
U.S. Government under contract No. DE-AC05-00OR22725.  Accordingly, the
U.S. Government retains a nonexclusive, royalty-free license to publish
or reproduce the published form of this
contribution, or allow others to do so, for U.S. Government
purposes.}
}
\author{
Xindong Wang and X.-G. Zhang}
\affiliation{Computer Science and Mathematics Division,
Oak Ridge National Laboratory, Oak Ridge, TN 37831-6164}

\date{\today}
\begin{abstract}
A general spin symmetry argument is proposed for spin currents in
semiconductors. In particular,
due to the symmetry with respect to spin polarization of
the helicity eigenstates of the Luttinger Hamiltonian for a
hole-doped semiconductor, the spin polarized flux
from a single helicity eigenstate induced by an external electric field, is
canceled exactly when all such contributions from eigenstates that are
degenerate in energy are summed.
Thus, the net spin current predicted by Murakami et al,
Science 301, 1348 (2003),
cannot be produced by such a Hamiltonian. Possible symmetry breaking mechanisms
which may generate a spin current are discussed.
\end{abstract}
\pacs{}
\maketitle
\newpage


Although generating and utilizing spin currents in metallic films and tunnel
junctions has become
a routine matter in spintronics research and application, generating spin
currents in semiconductors has attracted greater research efforts because of
the tremendous potential of combining spintronics with the well-developed
semiconductor technology. However,
lack of an effective method for spin injection into semiconductors has been the
main obstacle of progess.
The recent work by Murakami et al\cite{Shuichi} is remarkable that
it points an unexpected possible route for achieving spin injection. They
proposed that a spontaneous and dissipationless spin current would be produced
by an applied electric field in a hole-doped semiconductor, thus bypassing the
difficulty of traditional methods of spin injection. Other groups\cite{Sinova}
are considering alternative proofs for the existence of such a spin current.
The unconventional and extraordinary nature of this idea, and its potentially
wide impact in the field of condensed matter physics, calls for a careful and
skeptical scrutiny.
In this letter, we will examine the possibility of a spontaneous spin current
from the consideration of the spin symmetry of the electron wave function.
Symmetry rules are more general than the particular form of the Hamiltonian
one solves. Although our discussion will be focused on the Luttinger
Hamiltonian, for which a negative result is obtained, the approach is equally
applicable to other forms of
Hamiltonian that might be considered for spin transport.

The result of Murakami et al
was based on the consideration of the Luttinger Hamiltonian\cite{Luttinger},
\begin{equation}
H_0=\frac{\hbar^2}{2m}\left[\left(\gamma_1+\frac{5}{2}\gamma_2\right)k^2
-2\gamma_2({\bf k}\cdot{\bf S})^2\right],
\label{H0}
\end{equation}
which describes the valence bands for a large class of hole-doped
semiconductors. Note here that ${\bf S}$ represents the total angular momentum
of an electron. The expectation value of the spin part of the angular momentum
is related to ${\bf S}$ through $(1/3)\langle\psi|{\bf S}|\psi\rangle$.
Under the influence of a uniform electric field, they obtained
a result that a net spin current is induced in a direction perpendicular to
the field. This is a puzzling result because
the size of the spin current seems to be independent of the strength of
the coupling that is the origin of this spin current, the coefficient
$\gamma_2$ in Eq. (\ref{H0}). In fact, the spin current becomes a constant
when $\gamma_2$ approaches zero which seems to predict a spin
current in an isotropic free-electron gas.

The key to the solution in Ref. \cite{Shuichi} is the $4\times 4$ unitary
transformation $U({\bf k})$ which generates
the helicity eigenstates,
\begin{equation}
({\bf k}\cdot{\bf S})|{\bf k},\lambda\rangle = k\lambda|{\bf k},\lambda\rangle,
\end{equation}
by the diagonalization $U({\bf k})({\bf k}\cdot{\bf S})U({\bf k})^{\dagger}=
k{\cal S}_z$.  Here ${\cal S}_i$, $i=x,y,z$, are the $4\times 4$ spinor matrices for spin
$3/2$ (not to be
confused with the $i$th component of the spin matrix ${\bf S}$).
Although the choice of $U({\bf k})$ is not unique,
Ref. \cite{Shuichi} chose the simple form $U({\bf k})=\exp{(i\theta {\cal S}_y)}
\exp{(i\phi {\cal S}_z)}$. This represents a rotation in the spin space of angle $\phi$
about the $z$ axis followed by a rotation of angle $\theta$ about the $y$ axis.
Under this transformation, the coordinate system for the spin part of the
wave function has been rotated to a reference frame
which is local in the ${\bf k}$ space, and
the good quantum number is the helicity $\lambda$ within this frame.
The $\lambda$ eigenstate corresponds to a spin state that is precessing around
an axis parallel to ${\bf k}$, and
$\lambda$ is the projection of the spin along this axis.
The $\pm\lambda$ eigenstates at the same
${\bf k}$ point have exactly
opposite spin polarizations.
The spin symmetry of the helicity eigenstates are expressed as,
\begin{equation}
\langle{\bf k},\lambda|{\cal S}_i|{\bf k},\lambda\rangle =
-\langle{\bf k},-\lambda|{\cal S}_i|{\bf k},-\lambda\rangle,
\end{equation}
where $i=x,y,z$. 
A time-dependent spin operator, ${\bf S}(t)=
S_x(t)\hat{\bf x}+S_y(t)\hat{\bf y}+S_z(t)\hat{\bf z}$,
can always be expressed in the form with the $i$th component,
$S_i(t)=\sum_j \alpha_{ij}(t){\cal S}_j$, where $\alpha_{ij}(t)$ are elements
of a rotation matrix and are scalar
functions of time. Thus, the antisymmetry in $\lambda$
holds for any time-dependent spin operator,
\begin{equation}
\langle{\bf k},\lambda|{\bf S}(t)|{\bf k},\lambda\rangle =
-\langle{\bf k},-\lambda|{\bf S}(t)|{\bf k},-\lambda\rangle.
\label{S}
\end{equation}
Because the $\pm\lambda$ states are degenerate in
energy, the net spin from each ${\bf k}$ direction is exactly zero. This is
consistent with the fact that
semiconductors not doped with magnetic impurities generally do not
exhibit magnetism.
Similarly, using the following symmetrized three-spin product which
is also antisymmetric in $\lambda$,
\begin{equation}
\langle{\bf k},\lambda|\{{\cal S}_i,\{{\cal S}_j,{\cal S}_k\}\}|{\bf k},\lambda\rangle =
-\langle{\bf k},-\lambda|\{{\cal S}_i,\{{\cal S}_j,{\cal S}_k\}\}|{\bf k},-\lambda\rangle,
\end{equation}
with $i,j,k=x,y,z$, we find the symmetry relationship,
\begin{equation}
\langle{\bf k},\lambda|\{S_i(t),\{S_j(t),{\bf S}(t)\}\}
|{\bf k},\lambda\rangle =
-\langle{\bf k},-\lambda|\{S_i(t),\{S_j(t),{\bf S}(t)\}\}
|{\bf k},-\lambda\rangle,
\label{SSS}
\end{equation}
which will be needed in computing the spin current.

The question remains whether the spin polarization is canceled in a linear
response solution under a uniform electric field. Let us consider the transverse
component (perpendicular to the applied field)
of the linear-response current which arises from the first term on the rhs
of the equation of motion,
\begin{equation}
\dot{\tilde{x}_i}=\frac{\hbar k_i}{m_\lambda}+F_{ij}\dot{k}_j,
\label{xtdot}
\end{equation}
where $\tilde{x}_i=U({\bf k})x_iU^{\dagger}({\bf k})$ and we use a tilde to
clearly distinguish between the transformed operators $\tilde{x}_i$ and the
untransformed operators $x_i$. We will consider the meaning of the second
term later in this paper.
Karplus and Luttinger\cite{Karplus} pointed out in
their consideration of the anomalous Hall effect, that a transverse current
is the consequence of the spin-orbit coupling which introduces a left-right
asymmetry in the stationary (Bloch) states of the system.
The transverse current here similarly arises from the $({\bf k}\cdot {\bf S})^2$
term in the Hamiltonian.
What is remarkable about the transverse current predicted by Ref. \cite{Shuichi}
is that for each direction ${\bf k}$ the contribution to the
transverse current is completely independent
of the parameter $\gamma_2$, the coefficient of the $({\bf k}\cdot{\bf S})^2$
term in the Hamiltonian.
Thus it makes an unusual prediction of a finite spin
current in the direction perpendicular to the external electric field even
for an infinitesimal coupling constant $\gamma_2$. If this were indeed true,
it would cast doubt on the validity of the linear-response theory itself.
Fortunately, because the equation of motion is the same for both $\pm\lambda$
eigenstates, the transverse currents for both states are the same. Since
the spin polarizations of the two states are exactly the
opposite, there is no net spin current arising from this transverse
term. 

The total spin current can be calculated directly within the unrotated
global reference frame. The equations of motion without the rotation operations
are simply
\begin{equation}
\dot{\bf k}(t) = e{\bf E},
\label{kdot}
\end{equation}
and
\begin{equation}
\dot{\bf x}(t) = \frac{\hbar{\bf k}}{m}(\gamma_1+\frac{5}{2}\gamma_2)
-\frac{\hbar}{m}\gamma_2[({\bf k}\cdot{\bf S}){\bf S}+
{\bf S}({\bf k}\cdot{\bf S})].
\label{xdot}
\end{equation}
The electron flux at each ${\bf k}$ is
$\sum_{\lambda}\langle {\bf k},\lambda|\dot{\bf x}| {\bf k},\lambda\rangle$.
When summed over all ${\bf k}$, the total flux is cancelled except for the
drift contribution due to the combination of Eq. (\ref{kdot}) and scattering,
which within the linear response theory yields a term proportional to the
electric field ${\bf E}$.
If we set the $z$ direction as the direction of the applied electric field,
then the transverse spin flux can be calculated using the helicity
eigenstates, $|{\bf k},\lambda\rangle$, of
$({\bf k}\cdot{\bf S})$ which are
also the eigenstates of the Hamiltonian in Eq. (\ref{H0}).
The spin current is calculated from the linear response formula
${\rm Tr}[\rho \hat{\bf j}_{\rm spin}]$
where $\rho$ is the density matrix and the spin current operator is
given in the symmetrized form,
\begin{equation}
\hat{\bf j}_{\rm spin}=\frac{1}{2}\left\{{\bf S},\dot{\bf x}\right\}.
\label{jhat}
\end{equation}
This would be the typical starting point for deriving the Kubo-Greenwood
linear response formula.  However, in our case it is more straightforward
to use the Heisenberg representation, and directly apply the equations of
motion, Eqs. (\ref{kdot}) and (\ref{xdot}).
The flux in the $y$ direction with
the spin polarization along $x$ at each ${\bf k}$ point, is
\begin{eqnarray}
j_y^x({\bf k})&=&\frac{1}{6}\sum_{\lambda}\langle {\bf k},\lambda|
\{S_x(t),\dot{y}(t)\}| {\bf k},\lambda\rangle n^{\lambda}({\bf k})
\nonumber \\
&=&
\frac{\hbar k_y}{3m}(\gamma_1+\frac{5}{2}\gamma_2)
\sum_{\lambda}\langle{\bf k},\lambda|S_x(t)|
{\bf k},\lambda\rangle n^{\lambda}({\bf k})
\nonumber\\
&&-\frac{\hbar}{6m}\gamma_2{\bf k}\cdot \sum_{\lambda}
\langle {\bf k},\lambda|
\{S_x(t),\{S_y(t),{\bf S}(t)\}\}|{\bf k},\lambda\rangle n^{\lambda}({\bf k}),
\label{jxy}
\end{eqnarray}
where $S_x(t)$ and $S_y(t)$ are the components of the
spin matrix in the $x$ and $y$ directions, and $n^\lambda({\bf k})$
is the occupation number for the state $|{\bf k},\lambda\rangle$.
Before summing over $\lambda$, a linear response spin current would arise from
the drift term in
${\bf k}$. However, the spin factors can be shown to cancel exactly
at each ${\bf k}$.
Because of the degeneracy of the helicity eigenstates $|{\bf k},\lambda\rangle$
and $|{\bf k},-\lambda\rangle$, the occupation number,
$n^{\lambda}({\bf k}) = n^{-\lambda}({\bf k})$ is symmetric with respect to
$\lambda$.
Using the symmetry relations, Eqs. (\ref{S}) and (\ref{SSS}), we obtain,
\begin{equation}
j_y^x({\bf k})=0.
\end{equation}
Thus no transverse spin flux is induced by applying an electric field.

Now let us consider the second term in Eq. (\ref{xtdot}). We show here
that it results from the time derivative of the local reference frame.
First, let us rotate the equation of motion for ${\bf x}$, Eq. (\ref{xdot}),
to the local reference frame. In order to arrive at an expression comparable
to Eq. (\ref{xtdot}), one needs to
invoke the adiabatic approximation to drop the
off-diagonal terms coupling the light-hole (LH) and heavy-hole (HH) bands in
$U({\bf k})S_iU^{\dagger}({\bf k})= k_iS_z/k$.
We find,
\begin{equation}
U({\bf k})\dot{x}_iU^{\dagger}({\bf k}) = \frac{\hbar k_i}{m}
(\gamma_1+\frac{5}{2}\gamma_2-2S_z^2\gamma_2) = \frac{\hbar k_i}{m_{\lambda}}.
\end{equation}
This is exactly the first term in Eq. (\ref{xtdot}). It is clear, then, that
the time derivative of $\tilde{\bf x}$, includes two contributions,
\begin{equation}
\dot{\tilde{x}_i} = U({\bf k})\dot{x_i}U^{\dagger}({\bf k})+
\dot{\bf k}\cdot[\nabla_{\bf k}U({\bf k})x_iU^{\dagger}({\bf k})+
U({\bf k})x_i\nabla_{\bf k}U^{\dagger}({\bf k})],
\end{equation}
where the second term is equivalent to the second term in Eq. (\ref{xtdot})
and results from taking the time derivative of the rotation matrix $U$. This
clearly shows that $\dot{\bf x}$ and $\dot{\tilde{\bf x}}$ are not related
by a gauge transformation, the latter including a contribution from the
time derivative of the reference frame. Therefore it would be incorrect to
compute the spin current from $\dot{\tilde {\bf x}}$ as is done in
Ref \cite{Shuichi}. In fact, the validity of the adiabatic
approximation is also questionable.
However, this approximation is only needed to clearly separate
$\dot{\tilde{\bf x}}$ into two terms. It remains true that $\dot{\bf x}$ and
$\dot{\tilde{\bf x}}$ are not related by a gauge transformation whether or not
the adiabatic approximation is invoked.
We thus conclude that the nonzero spin current is an artifact of
their approach, not a physical solution.

A further comment on the generality of our result is in order. We have
shown that the cancellation of the spin current is due to the degeneracy
of the $\pm\lambda$ helicity eigenstates of the Luttinger Hamiltonian,
Eq. (\ref{H0}).
One is certainly tempted to consider the possibility that by breaking the
degeneracy between the $\pm\lambda$ states,
one might be able to produce a spin current from Eq. (\ref{jxy}).
For example, one might introduce an anisotropic kinetic energy term in Eq.
(\ref{H0}), such that it breaks the spatial inversion symmetry, thus
Eq. (\ref{jxy}) will need to include interband terms which are usually
not symmetric with respect to $\pm\lambda$. Another example that breaks the
symmetry
between the $\pm\lambda$ states has been considered by Ref.\cite{Sinova},
using the Rashba Hamiltonian which is linear in ${\bf k}$ and ${\bf S}$.
The negative result for the Luttinger Hamiltonian here does not rule out the
possibility of a spin current when such symmetry is broken.

In conclusion, we have shown that at least for a class of semiconductors
described by the Luttinger Hamiltonian, spin symmetry of the eigenstates
rules out the possibility of a spontaneous spin current in these materials.
Thus, any attempt to produce such a spin current must include symmetry
breaking terms in the Hamiltonian. This should provide important guidance to
future attempts of finding a mechanism for spin currents in semiconductors.


We thank Prof. Jairo Sinova for helpful discussions and in particular for
pointing out the problem with the adiabatic approximation
in Ref.\cite{Shuichi}.
This work was supported by the Office of Basic Energy
Sciences Division of Materials Sciences of the U.S. Department of Energy.
Oak Ridge National Laboratory is operated by UT-Battelle, LLC, for
the U.S. Department of Energy under contract DE-AC05-00OR22725.

\end{document}